\documentclass{ws-ijmpe}
\def\eV{\hbox{ eV}}

\begin{document}
%\wstoc
%{Suppression of $0\nu 2\beta$ Decay from CP Violation}
%{M. G\'o\'zd\'z, W. A. Kami\'nski}

\markboth
{M. G\'o\'zd\'z and W. A. Kami\'nski}
{Suppression of $0\nu 2\beta$ Decay from CP Violation}

%\catchline{}{}{}{}{}
%
%%%%%%%%%%%%%%%%%%%%%%%%%%%%%%%%%%%%%%%%%%%%%%%%%%%%%%%%%%%%%%%%%%%%

\title{SUPPRESSION OF $0\nu 2\beta$ DECAY FROM CP VIOLATION}

\author{MAREK G\'O\'ZD\'Z and WIES{\L}AW A. KAMI\'NSKI}

\address{
Department of Informatics, Maria Curie-Sk{\l}odowska University, \\
pl. Marii Curie--Sk{\l}odowskiej 5, 20-031 Lublin, Poland \\
mgozdz@kft.umcs.lublin.pl \\
kaminski@neuron.umcs.lublin.pl
}

\maketitle

%%%%%%%%%%%%%%%%%%%%%%%%%%%%%%%%%
\begin{history}                 %
\received{(received date)}      %
%\revised{(revised date)}       %
\accepted{(Day Month Year)}     %
%\comby{(xxxxxxxxxx)}           %
\end{history}                  	%
%%%%%%%%%%%%%%%%%%%%%%%%%%%%%%%%%

\begin{abstract}
  The observed phenomenon of neutrino oscillations is interpreted as the
  proof that neutrinos must have mass. As this is true for the neutrinos
  in the mass basis, the mass matrix in the flavor (weak) basis may
  still contain zeros. This can happen if the CP violating phases,
  usually neglected, come into play and result in suppression of
  processes which half-life depends on the masses of $\nu_e$, $\nu_\mu$,
  or $\nu_\tau$. In the present paper we investigate the possibility of
  such suppression of the neutrinoless double beta decay ($0\nu2\beta$).
\end{abstract}
%%%%%%%%%%%%%%%%%%%%%%%%%%%%%%%%%%%%%%%%%%%%%%%%%%%%%%%%%%%%%%%%%%%%

\section{Introduction}

The oscillations of neutrinos have been confirmed by a number of
experiments.\cite{mgozdz:nu-osc} The usual interpretation of this fact
within standard quantum mechanics implies that the quantities $\Delta
m_{ij}^2=|m_i^2 -m_j^2|$ must be non-zero, thus the neutrinos (with the
possible exception of the lightest one) must be massive. This calls for
need of extending the standard model of particles and interactions (SM),
in which neutrinos are assumed to be massless.

Neutrinos reveal also one more characteristic feature, namely, the
particles that take part in weak interactions, like the $\beta$ decay,
are not the same particles that propagate through space. In another
words, the $\nu_e$, $\nu_\mu$, and $\nu_\tau$ neutrinos, which are
produced in weak processes are said to be in the interaction (or flavor)
basis, while neutrinos that propagate (eg. from the Sun to Earth, from
radioactive source to the detector etc.) are in the physical basis and
are labeled by $\nu_1$, $\nu_2$, and $\nu_3$. The states $\nu_i$,
$i=1,2,3$, are mass eigenstates, so that one can speak about their
masses $m_i$. The weak eigenstates $\nu_\alpha$, $\alpha=e,\mu,\tau$,
are linear combinations of $\nu_i$,
\begin{equation}
  \nu_\alpha = U_{\alpha i} \nu_i,
\label{mgozdz:nu}
\end{equation}
and do not possess definite masses. The neutrino phenomenological mass
matrix in the flavor basis, ${\cal M}$, is usually written as
\begin{equation}
  {\cal M} = U^* {\rm diag}(m_1, m_2, m_3) U^\dagger,
\label{mgozdz:M}
\end{equation}
where $U$ in Eqs. (\ref{mgozdz:nu}) and (\ref{mgozdz:M}) is the unitary
Pontecorvo-Maki-Nakagawa-Sakata (PMNS) neutrino mixing matrix. For
massive Majorana neutrinos it can be conveniently parameterized by three
mixing angles $\theta_{12}$, $\theta_{13}$, and $\theta_{23}$, two
Majorana CP violating phases $\alpha_{12}$, and $\alpha_{13}$, and one
Dirac CP violating phase $\delta$:
\begin{eqnarray}
  U &=&
  \left (
    \begin{array}{ccc}
      c_{12} c_{13} & s_{12} c_{13} & s_{13} e^{-{\rm i} \delta} \\
      -s_{12} c_{23} - c_{12} s_{23} s_{13}e^{{\rm i} \delta} & c_{12} c_{23} - s_{12}
      s_{23} s_{13}e^{{\rm i} \delta} & s_{23} c_{13} \\
      s_{12} s_{23} - c_{12} c_{23} s_{13}e^{{\rm i} \delta} & -c_{12} s_{23} - s_{12}
      c_{23} s_{13}e^{{\rm i} \delta} & c_{23} c_{13}
    \end{array}
  \right ) \nonumber \\\nonumber \\
  &\times& {\rm diag} (1, e^{{\rm i} \alpha_{12}}, e^{{\rm i} \alpha_{13}} ),
\label{mgozdz:U}
\end{eqnarray}
where $s_{ij} \equiv \sin\theta_{ij}$, $c_{ij} \equiv \cos\theta_{ij}$.
Three mixing angles $\theta_{ij}$ vary between 0 and $\pi/2$. The CP
violating phases vary between $0$ and $2\pi$.

The investigation of neutrino oscillations provides pieces of
information about the mixing angles and the differences of masses
squared $\Delta m_{ij}^2$. It gives us, however, no information about
the absolute masses and the possible CP phases, which do not affect the
oscillation probabilities. To obtain the absolute values of neutrino
masses, one has to look for the neutrinoless double beta decay
($0\nu2\beta$),\cite{mgozdz:0n2b} investigate the end-point of the beta
decay spectrum in Tritium,\cite{mgozdz:katrin} or relay on the rather
rough cosmological models of supernova explosions, large scale structure
of the Universe etc.\cite{mgozdz:cosmo-nu} The aforementioned
$0\nu2\beta$ process is at present the only one known, capable of
distinguishing between Majorana and Dirac neutrinos, which makes the
searches for $0\nu2\beta$ particularly important.

Another problem, which is not resolved by the oscillation experiments,
is the hierarchy of masses. From the differences of masses squared one
cannot deduce the actual alignment of $m_i$. In agreement with the
present experimental knowledge are two scenarios:
\begin{romanlist}[(ii)]
\item the so-called normal hierarchy scenario (NH), which is defined by
  $$ m_1 \ll m_2 \ll m_3,$$ 
\item and the inverted hierarchy (IH), in which $m_3$ is the lightest one,
  $$ m_3 \ll m_1 < m_2.$$
\end{romanlist}
At present we have no hint which of the hierarchies is realized in
nature, so both of them have to be considered.

The neutrinoless double beta decay is a second order process, forbidden
in the standard model due to lepton number violation ($\Delta L=2$).  It
may occur only if the neutrinos are Majorana particles and some
mechanism of lepton number violation is introduced. These conditions are
fullfilled by many extensions of the SM, like the $R$-parity violating
MSSM and others. Therefore the $0\nu2\beta$ process, as the possible
source of information about the physics beyond SM, is intensively
investigated theoretically and searched for in experiments. The
half-life of $0\nu2\beta$ is proportional to the so-called effective
neutrino mass, which is just the ${\cal M}_{ee}$ element of the neutrino
mass matrix Eq. (\ref{mgozdz:M}). It is a common belief that the
confirmation of neutrino oscillations gives a strong back-up for the
$0\nu2\beta$ decay.  However, in such discussions one usually forgets
about the possible CP phases. In the present paper we show, that the CP
phases may in fact completely suppress the $0\nu2\beta$ process, while
still being in agreement with all the oscillation data.

\section{Calculations and Results}

The best-fit neutrino oscillation parameters may be summarized as
follows:\cite{mgozdz:nu-osc}
\begin{eqnarray}
  \Delta m^2_{23} = 2.1 \times 10^{-3}\eV^2, &\qquad& \sin^2 2\theta_{23}=1.00,
  \nonumber\\
  \Delta m^2_{12} = 7.1 \times 10^{-5}\eV^2, &\qquad& \tan^2 \theta_{12}=0.40.
  \nonumber
\end{eqnarray}
For the angle $\theta_{13}$ only the upper bound is known. From
exclusion plot obtained from the data of the reactor experiment
CHOOZ\cite{mgozdz:CHOOZ} we have
\begin{equation*}
  \sin^2\theta_{13} \le 0.05  \qquad \qquad (90\%\ \rm{c.l.}),
\end{equation*}
with zero being the best-fit value.

The elements of the neutrino mass matrix $\cal M$ become now a rather
complicated functions of the phases. Let us start with the case, in
which we take $\sin^2\theta_{13} = 0$. Now, directly from Eqs.
(\ref{mgozdz:M}) and (\ref{mgozdz:U}) one can obtain the following
expressions:
\begin{eqnarray}
  {\cal M}_{ee} &=&
  c_{12}^2 m_1 + s_{12}^2 m_2 e^{-{\rm i}2\alpha_{12}}
  \\
  {\cal M}_{e\mu} &=& -{\cal M}_{e\tau} = \frac{1}{\sqrt{2}} c_{12} s_{12}
  \left(-m_1+m_2 e^{-{\rm i} 2\alpha_{12}} \right),
  \\
  {\cal M}_{\mu\mu} &=& {\cal M}_{\tau\tau} = \frac12 \left(
    s_{12}^2 m_1 + c_{12}^2 m_2 e^{-{\rm i}2\alpha_{12}} + m_3 e^{-{\rm i}2\alpha_{13}}
  \right),
  \\
  {\cal M}_{\mu\tau} &=& -\frac12 \left(
    s_{12}^2 m_1 + c_{12}^2 m_2 e^{-{\rm i}2\alpha_{12}} - m_3 e^{-{\rm i}2\alpha_{13}}
  \right).
\end{eqnarray}
The elements ${\cal M}_{\alpha\beta}$ are complex so the physically
relevant quantities are $|{\cal M}_{\alpha\beta}|$. The strategy now is,
that we want to find such combination of the lightest neutrino mass
($m_1$ in NH and $m_3$ in IH) and CP phases to obtain $|{\cal
  M}_{\alpha\beta}|=0$. Because $|{\cal M}_{\alpha\beta}|= [\Re({\cal
  M}_{\alpha\beta})^2 + \Im({\cal M}_{\alpha\beta})^2]^{1/2}$, where
$\Re$ and $\Im$ stand for real and imaginary parts, respectively, we are
looking for solutions of the set of equations:
\begin{eqnarray}
  \left\{
    \begin{array}{l}
      \Re({\cal M}_{\alpha\beta}) = 0 \nonumber \\
      \Im({\cal M}_{\alpha\beta}) = 0
    \end{array}
  \right.
\end{eqnarray}
for each element in the case of NH and IH. 

Our results are presented in Tab. \ref{mgozdz:tab1}.
%
%%%%%%%%%%%%%%%%%%%%%%%%%%%%%%%%%%%%%%%%%%%%%%%%%%%%%%%%%%%%%%%%%%%%%%
% TAB 1 RESULTS %%%%%%%%%%%%%%%%%%%%%%%%%%%%%%%%%%%%%%%%%%%%%%%%%%%%%%
%%%%%%%%%%%%%%%%%%%%%%%%%%%%%%%%%%%%%%%%%%%%%%%%%%%%%%%%%%%%%%%%%%%%%%
\begin{table}[h!]
  \tbl{Results for the case $s_{13}=0$.\label{mgozdz:tab1}}{
  \begin{tabular}{ccc} 
    \toprule
    Element & Hierarchy & Mass \\ 
    \colrule
    $|{\cal M}_{ee}|=0$       & NH & $m_1 =   3.557 \times 10^{-3} \eV$ \\
    $|{\cal M}_{\mu\mu}|=0$   & IH & $m_3 \ge 2.271 \times 10^{-2} \eV$ \\
    $|{\cal M}_{\tau\tau}|=0$ & IH & $m_3 \ge 2.271 \times 10^{-2} \eV$ \\
    $|{\cal M}_{\mu\tau}|=0$  & IH & $m_3 \ge 2.272 \times 10^{-2} \eV$ \\
    \botrule
  \end{tabular}}
\end{table}
%%%%%%%%%%%%%%%%%%%%%%%%%%%%%%%%%%%%%%%%%%%%%%%%%%%%%%%%%%%%%%%%%%%%%%
% $|M_{ee}|=0$ for NH, $m_1 = 0.003556811647$, $\cos(2\alpha_{12})=-1$
% $|M_{\mu\mu}|=|M_{\tau\tau}|=0$ for IH, $m_3 \ge 0.022709914$
% $|M_{\mu\tau}|=0$ for IH, $m_3 \ge 0.02272416$
%%%%%%%%%%%%%%%%%%%%%%%%%%%%%%%%%%%%%%%%%%%%%%%%%%%%%%%%%%%%%%%%%%%%%%
%
As one sees, in the case of NH only the element ${\cal M}_{ee}$ may be
zero, and this situation requires a precisely fine-tuned value of $m_1$.
The corresponding phase $\alpha_{12}$ is given by the condition
$\cos(2\alpha_{12})=-1$. Other elements, of less interest for us, has
been calculated as well. It turned out that it is impossible to find
zero solutions for the $e\mu$ and $e\tau$ elements, while for the
remaining ones there are solutions in the case of IH only. One has to
bear in mind, that the lower bounds on $m_3$ come out from our
calculations but do not take into account other constraints on neutrino
masses. For example the global astrophysical and cosmological limit on
the sum of all three masses $\sum m_i$ is roughly given by 1~eV, which
means that $m_3$ cannot exceed this value. Notice also that, for obvious
reasons, it is impossible to have all diagonal elements being zero at
one time.

\begin{figure}[t]
  \centerline{\psfig{file=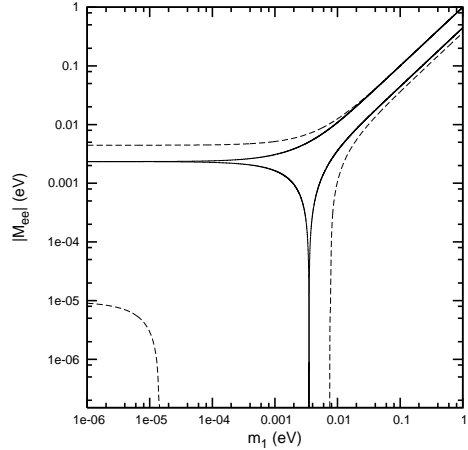,width=0.5\textwidth}}
  \vspace*{8pt}
  \caption{\label{mgozdz:fig1} $|{\cal M}_{ee}|$ as the function of
    lightest neutrino mass $m_1$ (normal hierarchy scenario). The dashed
    line corresponds to $\sin\theta_{13}=0.05$, the solid line to
    $\sin\theta_{13}=0$.}
\end{figure}

In the case in which $s_{13} \not = 0$, the expressions for ${\cal
  M}_{\alpha\beta}$ become very complicated functions.\cite{mgozdz:art9}
Some of them depend simultaneously on all three phases which makes their
analysis nearly impossible, also numerically. Fortunately, the $ee$
element remains relatively simple:
\begin{equation}
  {\cal M}_{ee} =
  c_{13}^2 c_{12}^2 m_1 + c_{13}^2 s_{12}^2 m_2 e^{-{\rm i}2\phi_2} 
  + s_{13}^2 m_3 e^{-{\rm i}2(\alpha_{13}-\delta)}. 
\end{equation}
Assuming the maximal allowed value $s_{13}^2 = 0.05$ one obtains the
zero solution for NH and 
\begin{equation*}
  8.047\times 10^{-5} \eV \le m_1 \le 7.473\times 10^{-3} \eV. 
\end{equation*}
One notices that, as expected, this solution contains the one obtained
in the simpler case $s_{13}=0$. The possible values of $|{\cal M}_{ee}|$
are depicted on Fig.~\ref{mgozdz:fig1}.

%%%%%%%%%%%%%%%%%%%%%%%%%%%%%%%%%%%%%%%%%%%%%%%%%%%%%%%%%%%%%%%%%%%%%%
% FIG. 1 %%%%%%%%%%%%%%%%%%%%%%%%%%%%%%%%%%%%%%%%%%%%%%%%%%%%%%%%%%%%%
%%%%%%%%%%%%%%%%%%%%%%%%%%%%%%%%%%%%%%%%%%%%%%%%%%%%%%%%%%%%%%%%%%%%%%
%%%%%%%%%%%%%%%%%%%%%%%%%%%%%%%%%%%%%%%%%%%%%%%%%%%%%%%%%%%%%%%%%%%%%%

\section{Conclusions}

We have shown that the popular practice of neglecting CP phases in the
neutrino mass matrix may have severe consequences. In particular, for
certain combination of parameters (phases and the mass of the lightest
neutrino), weak processes like the neutrinoless double beta decay may be
suppressed. However, the inclusion of CP phases in the calculations is a
highly non-linear problem, which may be a big obstacle. It is thus
desirable to find a method of determining the possible CP violation in
the neutrino sector experimentally.

It is interesting to note, that the obtained values of $m_1$ which give
zero solutions for the ${\cal M}_{ee}$ element are within the range of
experimental and theoretical bounds (roughly speaking they do not exceed
the limit of 1 eV).

\section*{Acknowledgements} 

The first author is grateful for the financial support from the
Foundation for Polish Science. He would like also to express his
gratitude to prof. A. Faessler for his hospitality at the University of
T\"ubingen during the Summer 2006.

%%%%%%%%%%%%%%%%%%%%%%%%%%%%%%%%%%%%%%%%%%%%%%%%%%%%%%%%%%%%%%%%%%%%%%

%%%%%%%%%%%%%%%%%%%%%%%%%%%%%%%%%%%%%%%%%%%%%%%%%%%%%%%%%%%%%%%%%%%%%%

\end{document}